\documentclass[preprint2]{aastex}
\usepackage{graphicx}
\usepackage{color}
\usepackage{textcomp}
\usepackage{amssymb}

\shorttitle{Mono-Metallicity Populations as evidence for Two-Phase Disk Formation}
\shortauthors{Dom\'{\i}nguez-Tenreiro et al.}

\begin{document}

\title{The Radial Distribution of  Mono-Metallicity Populations in the Galactic Disk as Evidence for Two-Phase Disk Formation  }

\author{R. Dom\'{\i}nguez-Tenreiro\altaffilmark{1,2}, A. Obreja\altaffilmark{1,3}, C.B. Brook\altaffilmark{1,2,4},
F.~J. Mart\'{\i}nez-Serrano \altaffilmark{5,6} \& A. Serna\altaffilmark{5}}

\altaffiltext{1}{Dept. de F\'{i}sica Te\'orica, Univ. Aut\'onoma de Madrid, E-28049  Cantoblanco Madrid, Spain}
\altaffiltext{2}{Astro-UAM, UAM, Unidad Asociada CSIC, E-28049 Cantoblanco, Madrid, Spain} 
\altaffiltext{3}{University Observatory Munich, Scheinerstr. 1, D-81679 Munich, Germany}
\altaffiltext{4}{Instituto de Astrof\'{i}sica de Canarias, Univ. de La Laguna, E-38206 La Laguna, Tenerife, Spain}
\altaffiltext{5}{Dept. de F\'{i}sica y A.C., Universidad Miguel Hern\'{a}ndez, E-03202 Elche, Spain}
\altaffiltext{6}{present address: Next Limit Dynamics SL,
      Angel Cavero 2, E-28048  Madrid, Spain}
\email{rosa.dominguez@uam.es}

\begin{abstract}

Recent determinations of the radial distributions of mono-metallicity populations (MMPs,
 i.e., stars in narrow bins in [Fe/H] within wider [$\alpha$/Fe] ranges) 
by the  SDSS-III/APOGEE DR12 survey cast  doubts on the classical thin - thick disk dichotomy.
The analysis of these observations lead to the non-$[\alpha$/Fe] enhanced populations splitting 
into MMPs with different surface densities according to their [Fe/H].
 By contrast,  $[\alpha$/Fe] enhanced (i.e., old) populations show an homogeneous behaviour.
We analyze these results in the wider context of disk formation within 
non-isolated halos embedded in the Cosmic Web, resulting in a two-phase mass assembly. 
By performing hydrodynamical simulations in the context of the $\rm \Lambda CDM$  model, 
we have found that the two phases of halo  mass assembly 
(an early, fast phase, followed by a slow one, with low mass assembly rates)
are very relevant to determine
the radial structure of  MMP distributions, while radial mixing has only a secondary role, 
depending on the coeval dynamical and/or destabilizing events. 
Indeed, while the frequent dynamical violent events occuring
at high redshift remove metallicity gradients, and  imply efficient stellar mixing, 
the relatively quiescent dynamics
after the transition keeps [Fe/H] gaseous gradients and prevents newly formed stars to suffer
from strong radial mixing. 
By linking the two-component disk concept with the two-phase halo mass assembly
scenario, our results set halo virialization (the event marking the transition from the
fast to the slow phases) as the separating event marking periods characterized by
different physical conditions under which thick and thin disk stars were born.

\end{abstract}

\keywords{cosmology: theory, galaxies: formation, methods: numerical}

\section{Introduction}
\label{Intro}

Recently, the spectra of some 70,000 red giant stars from SDSS-III/APOGEE DR12 \citep{Majewski:2015} have been
obtained in the H-band, where the dust effects are not  important, 
providing the element chartography of the Milky Way (MW) over an unprecedented large volume
and including for the first time the Galactic plane. 
This very recent data opened up the possibility to study in more detail the processes 
occuring along the Milky Way (MW) assembly and evolution, through the imprints they have left into the stellar space distributions.

It has long been known that the stellar populations of spiral galaxies
have two main components: a dynamically hot spheroid 
and a cold disk. 
A third component, originally detected in the Milky Way through
stellar counts \citep{GilmoreReid:1983}, has  shown up as an ubiquitous excess of red flux
at large galactic latitudes in external spiral galaxies \citep{Dalcanton:2002,Yoachim:2006}. Detailed kinematic and chemical
studies \citep[e.g.][]{Fuhrmann:1998,Bensby:2003,Soubiran:2003,BensbyFeltzing:2005,Reddy:2006}
confirmed  that such excess was due to  a  distinct component, the thick disk.
Later on, statistics studies of thick disks in the local Universe suggests that they are  ubiquous in galaxies
\citep{Comeron:2011}. 

Observations 
\citep{Gilmore:1989,Gilmore:1995,Fuhrmann:1998,Bensby:2003,Soubiran:2003,Reddy:2006,Ivezic:2012}
indicate that the differences between the thin and thick disks involve:
i) shape: the vertical scale-length is smaller for the thin disk than for the thick one \citep[e.g.][]{Dalcanton:2002, Yoachim:2006},
ii) kinematics: the thin disk is colder in all velocity components, and
it is more rotationally supported than the thick disk \citep{Soubiran:2003},
iii) age: thick disk stars are older on average \citep{Gilmore:1995},
iv) metallicity: thin disk stars are more metal rich than thick disk ones \citep{Fuhrmann:1998}, and
v) $[\alpha$/Fe]: the thick disk has enhanced $\rm\alpha$-elements
compared to thin disk populations of similar [Fe/H] abundances
\citep[e.g.][]{Fuhrmann:1998,Bensby:2003,Reddy:2006},
 suggesting shorter star formation time-scales
 \citep{Fuhrmann:1998,Ruchti:2010}.

These findings have been confirmed by  observational studies made within recent or on-going spectroscopic surveys, such as:
the RAdial Velocity Experiment \citep[RAVE,][]{Steinmetz:2006},
the Sloan Extension for Galactic Understanding and Exploration  \citep[SEGUE,][]{Yanny:2009},
Gaia-ESO \citep{Gilmore:2012},
the Large Sky Area Multi-Object Fiber Spectroscopic Telescope  \citep[LAMOST,][]{Cui:2012},
Apache Point Observatory Galactic Evolution Experiment \citep[APOGEE,][]{Majewski:2015},
Gaia \citep{Brown:2016}.
The full explotation of these, as well as other planned surveys, 
such as the  GALactic Archeology with Hermes \citep[GALAH,][see also Duong et al. 2017, to be published]{Martell:2017}, 
or WEAVE, make of this decade a Golden Age for Astroarcheology, where
a dramatic advancement in the understanding of the Galaxy  
is expected.

While authors agree on these distinct properties of thin and thick disks,  
different methods to assign a given star to the thin or thick populations can be found in literature.
A kinematic  classification scheme   
is used extensively in hydrodynamical simulations
\citep[e.g.][]{Abadi:2003, Domenech:2012, DominguezTenreiro:2015}, and
 in observations \citep{Soubiran:2003, Bensby:2003,Kordopatis:2011,Boeche:2013a}.
Other classifications rest on  chemical properties such as the [$\rm\alpha$/Fe] vs [Fe/H] 
\citep[e.g.,][and references therein]{Bovy:2012,Adibekyan:2012,Ramirez:2013,RecioBlanco:2014,Bensby:2014,Hayden:2015,Martell:2017},
 or stellar age \citep[see for example][]{Kubryk:2015}.
\citet{Haywood:2013}
use a classification based on the [Si/Fe]
vs age relation, where the  separation comes from a  knee  in the two-slope behaviour of this relation
that these authors explain in terms of a particular shape for the star formation rate history 
 (SFRH), see \citet{Snaith:2014}. 
The two-slope behaviour for [$\alpha$/Fe] - age has been extended to local,
 early type galaxies by
\citet{Walcher:2015}.


A lot of effort has been so far devoted to understand the origin of the thick disk.
Basically two different scenarios exist  to explain the thick disk emergence 
\citep[see reviews in][]{Freeman:1987,Gilmore:1989,Freeman:2002,Brook:2004,vdKruit:2011,Ivezic:2012,Feltzing:2013,MinchevCharla:2017}.
The first one links thick disks
to  violent formation processes from turbulent gas prior to thin disk formation.
The second scenario assumes 
a preexisting thin disk that is dynamically heated along secular evolution, with stellar migration 
\citep[first proposed by][]{Roskar:2008,Schonrich:2009b,Loebman:2011} or stellar satellite acretion \citep{Abadi:2003} as the main paths.

The first scenarios are supported  in particular from cosmological simulations
 by \citet{Brook:2004,Brookb:2012}, \citet{Stinson:2013b},
 and  \citet{Bird:2013}, see also \citet{Scannapieco:2011}.
\citet{Brook:2004} identified a period of fast merging at high $z$ where chemically classified thick disk stars form kinematically hot,
 prior to the formation of thin disk stars.
 \citet{Brookb:2012} confirm the previous scenario and add radial mixing to explain
thick disk radial changes. They find that radial mixing has comparatively  a minor role in disk thickening and heating.
\citet{Stinson:2013b} extended these results and incorporate the metallicity disk structure to their analysis.

\citet{Scannapieco:2011} find that young (old) stars define thin (thick) disk structures, with kinematic and chemical properties reminescent of those of observed thin (thick) disks.

 \citet{Bird:2013} analyze the radial profiles of surface mass density and kinematics of mono-age stellar
populations, aided by a kinematical separation of their simulated galaxy components. They find that most of the kinematically defined
thick disk stars form within the first 4 Gyr after the Big Bang, in a dynamically violent period.
Their (kinematically defined)  thin disk stars formed that early
would be members of a chemically thick disk population, recovering in this way \citet{Brookb:2012}  results that thick disk stars are
born early and hot, while thin disk stars are not. They also agree on the role  radial mixing has  on disk heating and thickening.

 The ability of radially migrated inner stars to cause
thick, dyanmically hot extended disks is carefully analyzed in \citet{Minchev:2012},
using both cosmological simulations and pre-prepared ones.
In both cases, they also conclude that this process does
little for disk heating and thickening, but can change the spatial chemical structure of the disks
as shown by \citet{Kubryk:2013}.
These changes, however, do not wash out the chemical imprints left by disk formation according to the pre-prepared simulation 
results of \citet{Curir:2014}. 
Similar results on the effects of radial migration along  secular evolution have been presented by \citet{VeraCiro:2014}.

Some observational results suggest or are consistent with the previous scenario.
For example, \citet{Haywood:2013}, \citet{Snaith:2014} and \citet{Haywood:2015},
who, from an analysis of the chemical properties of a set of solar vicinity FKG stars \citep{Adibekyan:2012}, combined with a careful determination of their age, 
conclude that the thick disk stars have formed at early times out of a turbulent gas, setting the
chemical conditions for a latter thin disk formation in a more quiescent  situation.

 Different results from different surveys point in this direction. 
For example, \citet{Kordopatis:2015}    
found, in their analyses of data from the Gaia-ESO survey, that the mixing of metals in the young Galaxy
(e.g. turbulent gas in disks or radial stellar migration) was more efficient at early times before the (current) 
thin disk started forming \citep[see also][]{Mikolaitis:2014}.
Other authors also prefer this scenario on the basis of their results on  radial and/or vertical abundance gradientes from
different surveys \citep[see, for example][from RAVE and LAMOST data, respectively]{Cheng:2012,Xiang:2015} or consider
this possibility \citep{Boeche:2014}.
\citet{Comeron:2015} present the first Integral Field Unit spectroscopy of an edge-on galaxy, ESO 533-4,
 with enough depth and quality to study the thick disk. Even if not conclusive, their results suggest that the thick disk of ESO 533-4 formed in a relatively short event prior to the thin disk.

As for the second scenario, authors highlight different processes acting on a preexisting thin disk as the origin 
of the thick disk: radial migration \citep{Roskar:2008,Schonrich:2009b,Loebman:2011},
stars accreted through satellites and  dragged into the plane of a pre-existing disk \citep{Abadi:2003,Meza:2005}, heating of a
pre-existing disk by satellite accretion \citep{Quinn:1993} and later on confirmed by other authors \citep[see, e.g.][]{Hayashi:2006,
Kazantzidis:2008,Villalobos:2008,Bekki:2011,Qu:2011} and more recently by \citet{Moetazedian:2016} and \citet{RuizLara:2016}. 
The secular thickening of a disk embedded in a fluctuating potential has also been investigated  \citep{Fouvry:2016,Pichon:2016},
as well as the effects of early massive clump formation in unstable gas-rich disks 
 \citep[e.g.][]{Noguchi:1999,Bournaud:2007,Agertz:2009,Ceverino:2010},
see however \citet{Buck:2017}
or the popping of stellar clusters \citep{Kroupa:2002,Assmann:2011} on disk thickening. These  two last scenarios link thickening to formation
processes rather than to the heating of a pre-existing thin disk.

We see that there is currently a living debate  on the origin of the thick disk. The metallicity cartography of the Galactic disk provided
by the APOGEE project contribute new very interesting  possibilities therein.


Of particular interest for studies on the thin and thick disk are 
mono-metallicity populations
(MMPs, i.e., stars in narrow bins in [Fe/H] within wider [$\rm\alpha$/Fe] ranges). 
Analyzing the red giant sample from SDSS-III/APOGEE DR12, 
\citet{Bovy:2015} present a careful determination of their radial  
structure, and show that radial MMPs distributions
 show a bimodal behaviour according to their [$\rm\alpha$/Fe] content:  
i) MMPs with enhanced [$\rm\alpha$/Fe] show disk radial surface densities, $\rm\Sigma$(R$_{\rm cyl}$), that are well described by 
single exponential distributions, with a unique scale length 
no matter the MMP, confirming, with a much better statistics and radial coverage,
 previuos results by  \citet{Bovy:2012c} in the SDSS/SEGUE
survey, see also \citet{Kordopatis:2012}.   
ii) MMPs with low-[$\rm\alpha$/Fe] show a more
complex behaviour, with their respective MMP radial surface densities exhibiting continuously varying shapes, 
more centrally concentrated as [Fe/H] increases, 
and whose overall envelope is an exponential disk.

Taking enhanced and low-[$\alpha$/Fe] stellar populations as  thick and thin disk populations, respectively,   
\citet{Bovy:2015} maintain that  the results above 
cast  doubts on the classical thin - thick disk dichotomy,
because of the complexity of thin disk splitting into MMPs with different surface densities.
This issue has not yet received a complete answer so far \citep[but see][]{Haywood:2016,Minchev:2017},
 in particular when a cosmological context is envisaged. 
The aim of this paper is to  analyze this question 
within the wider context  of disk formation inside non-isolated halos, more particularly
 whithin the two-phase assembly scenario of halo formation. 
Analytical models as well as N-body simulations show that two different phases can be distinguished along the
halo mass assembly, as first proposed by \citet{Wechsler:2002,Zhao:2003,Brook:2005,SalvadorSole:2005},
 see also \citet{Griffen:2016} for recent results. These are:
i) first a violent rapid phase with high mass aggregation rates, resulting from 
collapse-like events in the Cosmic Web environment, implying high merger rates,
 and ii) later on a slow phase with lower mass aggregation rates.
Small box hydrodynamical simulations \citep{DT:2006} as well as larger box ones
\citep{Oser:2010,DT:2011} confirmed this scenario,
as well  as its implications on a possible scenario for thick disk formation \citep{Brook:2004},
elliptical properties at low redshift  \citep{Cook:2009}, and
on classical bulges \citep{Obreja:2013}. 

In this paper  we employ hydrodynamical simulations in a cosmological context in order to  address  the question:
 to what extent can the  
two-phase halo assembly scenario determine the  detailed MMP disk chartography, now available, in terms of
the thin versus thick disk differentiation ?
To answer to this question, we need to find out the physical basis of the thin versus thick disk 
differentiation within the two-phases scenario, an issue addressed in depth in this work. 

The paper is organized as follows. 
In Section \ref{MetRes} we describe the codes and the simulated galaxies.
The occurrence of a two-phase mass assembly for these galaxies
is examined in $\S$\ref{TwoPhase}, where the separating event between the phases is identified as halo virialization, 
defining a timescale t$_{\rm vir}$. 
In this section we propose a classification scheme for stars into thick and thin populations based on t$_{\rm vir}$
and in $\S$\ref{tchem} we study its relationship with a chemical classification based on age. 
In $\S$\ref{ThinThickProp} we confirm that this classificaction scheme leads
indeed to thin and thick disk populations whose properties differ in agreement with observational data.
The radial MMP distributions for the disks of simulated galaxies are analyzed in $\S$\ref{MMPsimu}, where we show that   
\citet{Bovy:2015} results are recovered. To decipher the physical processes underlying this behaviour, the 
stellar birth places are linked to their birth times and metallicity in $\S$\ref{BirthGasStr}, and are analyzed 
in terms of the gas metallicity structure before and after t$_{\rm vir}$. 
With these findings in mind, we come back to the two-phase galaxy assembly in $\S$\ref{TwoPhaseAgain}, 
and analyze the effects each phase has  on the physical conditions for star formation, 
gas metallicity gradients maintenance and removal, 
and the matching of the MMP distributions at stellar birth to those observed at z=0. 
The summary, discussion and conclusions are presented in $\S$\ref{SummDisConc}.

\section{Codes and Simulations}
\label{MetRes}
Accurate  conservation of angular momentum, a detailed implementation
of chemical evolution, and the effects of discrete energy injection by stellar physics  are  fundamental issues in cosmological hydrodynamical simulations. 
Comparisons of results using different codes is  advisable, because
we are looking for effects coming from a generic and fundamental level of physical description.   
In view of these considerations, in this paper we present results of  simulations run
with two different SPH codes: {\tt P-DEVA} and {\tt GASOLINE}. 

{\tt P-DEVA}  \citep{MartinezSerrano:2008} is an entropy conserving AP3M-SPH code
where the main concern in its design was angular momentum conservation.
Chemical evolution implementation
makes use of the Q$_{\rm ij}$ formalism \citep[][]{Talbot:1973}, 
which relates each nucleosynthetic product to all its different  sources. 
Ejecta from SNe as well as from low and intermediate mass stars have been taken into account.
The stellar evolution data described in \citet{Gavilan:2005} have been used for low and intermediate mass stars,
 and those in \citet{Woosley:1995} for high-mass stars. 
SNIa rates were computed according to \citet{RuizLapuente:2000} and their element production according to 
\citet{Iwamoto:1999}. We have considered the evolution of the following elements: H, He$^4$, C$^{12}$, C$^{13}$, N$^{14}$, O$^{16}$,
 Ne$^{20}$, Mg$^{24}$, Si$^{28}$, S$^{32}$, Ca$^{40}$, and Fe$^{56}$.
Stellar feedback is implicitly implemented through  (inefficient) star formation parameters,
as discussed in \citet{Agertz:2011}.

In {\tt GASOLINE} \citep{Wadsley:2004} SNe feedback is implemented using the 
blastwave-formalism \citep{Stinson:2006}. In the MaGGIC runs \citep{Brooka:2012} 
feedback from massive stars is also taken into account \citep{Stinson:2013a}.
The chemical evolution implementation  follows \citet{Raiteri:1996}, and we track 
nine elements: H, He$^{4}$, C$^{12}$, N$^{14}$, O$^{16}$, 
 Ne$^{20}$, Mg$^{24}$, Si$^{28}$, and Fe$^{56}$.
Metals are ejected from SNII, SNIa, and the stellar winds driven from asymptotic giant branch (AGB) stars,
using literature yields for SNIa \citep{Nomoto:1997}. Each SN releases 10$^{\rm 51}$ erg, 
and the cooling is delayed for 4 Myr for the neighbouring gas particles.
Ejected mass, metals and thermal energy from SNe are distributed to the nearest neighbour gas particles using the smoothing kernel 
\citep{Stinson:2006}. 

In both codes  the star formation recipe follows a Kennincutt – Schmidt- like law with a given
 density threshold, $\rm\rho_{\rm *}$, and star formation efficiency c$_{\rm *}$.
In both codes it is assumed that stars more massive than                                
8 M$_{\rm\odot}$ produce type II supernovae (SNII), and  the yields of \citet{Woosley:1995}, 
with the Fe ejecta divided by two according to \citet{Timmes:1995}, have been adopted. Also 
stars  take and retain their progenitor gas particle abundances.
The abundance increments synthetyzed by a given stellar particle within an integration timestep
are transferred to its nearest gas particle allowing stars to share their yields with the surrounding interstellar medium.

In both codes, abundance diffusion within the gas is included  based on (unresolved) turbulent mixing
\citep[see, for example,][]{Scalo:2004}, ensuring that 
gas abundance increment reaches farther away from the stellar position and explosion sites.
This metal mixing tends to equilibrate the element content of spatially close gas particles, 


\begin{table*}
\begin{minipage}{6.3in}
\renewcommand{\thefootnote}{\thempfootnote}
\begin{center}
\begin{tabular}{|c|c|c|c|c|c|c|c|c|c|c|} \hline
Object & $\rm\delta$M$_{\rm bar}$ & h$_{\rm soft}$ & $\rm\rho_{\rm *}$ & c$_{\rm *}$ &IMF & M$_{\rm *}$ & M$_{\rm gas}$ & t$_{\rm chem}$  &L$_{\rm box}$\\
 & 10$^{\rm 5}$M$_{\rm\odot}$ & h$^{\rm -1}$kpc & cm$^{\rm -3}$ & \% &&  10$^{\rm 10}$M$_{\rm\odot}$ & 10$^{\rm 10}$M$_{\rm\odot}$& Gyr & Mpc\\ \hline\hline

g1536-L  & 1.90 & 0.15 & 9.4 & 3.3 &Chab03& 2.32 & 1.97 & 8.0 & 34\\
HD-5004A & 3.94 & 0.20 & 6.0 & 1.0 &Salp55&  3.26 & 0.67 & 7.7 & 10\\
HD-5101A & 3.79 & 0.20 & 12.0 & 0.8 &Salp55&  1.29 & 0.33 & 7.2 & 10\\ \hline
\label{tab1}
\end{tabular}
\caption{The initial mass of gas particles ($\rm\delta$M$_{\rm bar}$, it is fixed for {\tt P-DEVA} runs), 
minimum SPH smoothing length (h$_{\rm soft}$), density threshold ($\rm\rho_{\rm *}$), 
star formation efficiency (c$_{\rm *}$), initial mass function (IMF), stellar (M$_{\rm *}$) and gas (M$_{\rm gas}$) 
mass of the simulated galaxies  at z=0, Universe age used to separate the thick from the thin stellar populations (t$_{\rm chem}$), 
and periodic box length (L$_{\rm box}$).}

\end{center}
\end{minipage}
\end{table*}



\begin{figure*}
\includegraphics[width=0.99\textwidth]{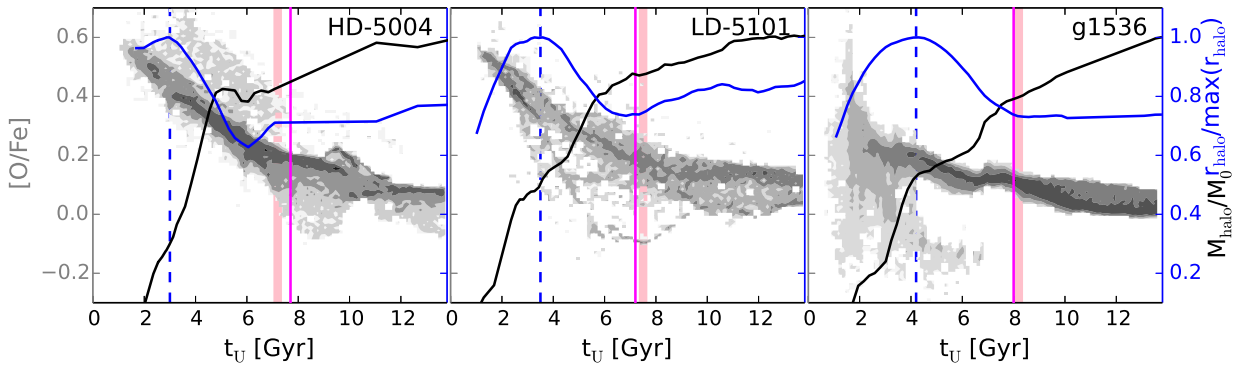}
\caption{Evolution of the virial mass M$_{\rm halo}$(t$_{\rm U}$) normalized to its z=0 value (black curves)
and of the radius r$_{\rm halo}$(t$_{\rm U}$) enclosing the particles belonging to the z=0 halo,
 normalized to its maximum value (blue curves), for the three simulated galaxies.
The redshifts of turn-around or maximum expansion, z$_{\rm turn}$, have been identified
as that corresponding to the maximum of the r$_{\rm halo}$(t$_{\rm U}$) curve,
and are shown as the dashed blue lines. 
We also mark the halo virialization time interval t$_{\rm vir}$ as pink bands. 
The corresponding [O/Fe] vs t$_{\rm U}$ relation for the stars in the disk of each galaxy are plotted in grey, 
with the intensity corresponding to the logarithm of the number of particles at each pixel in the plane.  
The magenta vertical lines mark the t$_{\rm chem}$ positions, separating thick disk stars (left) from thin disk ones (right). }

\label{Vir-MAT}
\end{figure*}



\begin{figure*}
\centering 
\includegraphics[width=0.85\textwidth]{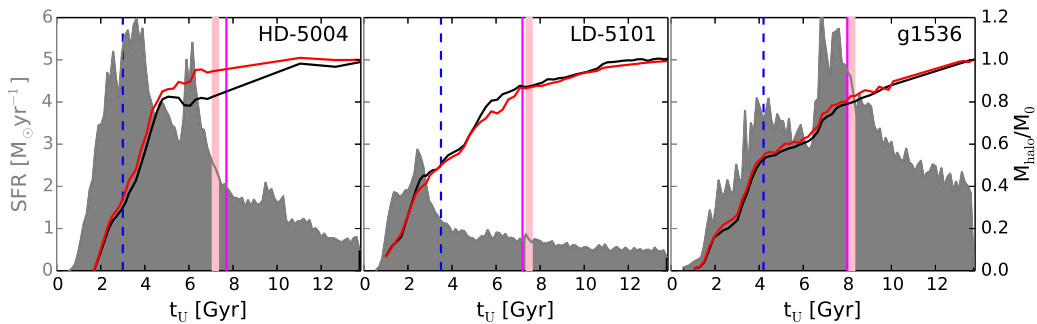}
 \caption{The star formation rate histories for the three simulated galaxies are plot in grey.
The blue and magenta vertical lines, and the pink bands, have the same meaning as in Figure 1.
Black lines are M$_{\rm halo}$(t$_{\rm U}$). 
Red lines are the baryon  
masses of the galaxy at each time t$_{\rm U}$.
}
   \label{SFRH}
\end{figure*}



A zoom-in simulation technique  has been used, with initial baryon particle masses $\rm\delta$M$_{\rm bar}$,
 minimal smoothing length  h$_{\rm soft}$, and periodic box length L$_{\rm box}$ given in Table 1, where we
also give the star formation efficiency c$_*$, and density threshold $\rm\rho_*$ parameters.
We explore two P-DEVA galaxies (HD-5004A and LD-5101A)
and one GASOLINE - MaGICC galaxy  (g1536-L$^*$). Their  stellar and gaseous  masses are given in Table 1.
The main difference between the two
P-DEVA galaxies is that, by construction,  HD-5004A forms and evolves in a dense environment,
while LD-5101A lives in a lower density one.
For the g1536-L$^{\rm *}$ galaxy the halo to be zoomed-in has been chosen to have  low merging activity.

These P-DEVA galaxies have been studied by
\citet{Domenech:2012} at z=0,
who  analyze some of their thin versus thick population properties,
GASOLINE ones by \citet{Brookb:2012}, \citet{Stinson:2013b} and \citet{Obreja:2014}.
In addition, \citet{Obreja:2013}
who addressed bulge formation, 
\citet{Dominguez:2014} who analyzed their HI and H2 content and panchromatic SEDs,
and \citet{Miranda:2016} who studied abundance gradients,
included galaxies run using both codes.  
In all these cases 
the consistency with observational data is very satisfactory, 
including the differences between 
the thin and thick disk described in $\S$\ref{Intro} analyzed in these papers.

\section{A two-phase mass assembly ?}
\label{TwoPhase}

A key issue in this paper is to elucidate if the halos of the analyzed systems had their mass assembled
through a two-phase process, as argued in $\S$\ref{Intro}.

The answers lie in Figure \ref{Vir-MAT}, where the evolution of the halo mass assembly  is illustrated through
the plot of the  halo virial mass, M$_{\rm halo}$(t$_{\rm U}$), as a function of time (black curves).
In this figure mergers appear as  discontinuities, $\rm\Delta$M$_{\rm halo}$(t$_{\rm U}$), spanning their respective
 merger time interval.
Expectedly (see $\S$\ref{Intro}),
we see that two phases clearly stand out, a fast one, where
the rate of mass assembly (i.e., the curve  slope at given times)  is very fast (high slopes),
and a slow one, when this rate is low or very low (low slopes),
see, i.e. \citet{Griffen:2016} for a recent similar result.
Note also that, at high redshifts,  mergers have their
respective merger time intervals superimposed in some cases (resulting in a rather continuous slope),
 with $\rm\Delta$M$_{\rm halo}$(t$_{\rm U}$) showing  a wide range of values. 
In the slow phase the discontinuities are less frequent and tend to be smaller, 
with no major mergers in any of the three galaxies.
An analysis of these curves in detail, taking into account the cosmological context, indicates that the frequent mergers
at early times are but the effects of the collapse of the region surrounding the halo\footnote{Therefore collapse and high redshift mergers have similar meaning.}.
This explains the violence of the mass-building events at high redshift, and is the basis for the so-called two-phase formation scenario.

As a second proof of the two-phase behaviour shown by haloes, 
in each of the three panels of  Figure  \ref{Vir-MAT} we plot 
the radii, r$_{\rm halo}$(t$_{\rm U}$), 
that enclose the particles forming the respective z=0  haloes at different t$_{\rm U}$ (blue curves). 
The radii have been normalized to their respective maximum values.
We see that these radii first increase, until they reach a maximum (turn-around) and then they decrease,
until they come to a quasi-equilibrium value retained up to z=0.
This r$_{\rm halo}$(t$_{\rm U}$) behaviour indicates that at high redshift, after expansion and turn-around, we witness halo collapse and  virialization   
in consistency with the spherical collapse model predictions, see \citet{Padma:1993}, followed by a slower period of evolution.
 In each panel of this figure
 the short time interval separating the decreasing  from the quasi-equilibrium  behaviour of r$_{\rm halo}$(t$_{\rm U}$) curve
(marked with a pink band) is close to, and in some cases rather coinciding with,
the short time interval separating the  halo mass assembly rates from fast to slow.
This is expected because  both dynamical timescales mark halo virialization. Therefore, 
only a dynamical timescale  t$_{\rm vir}$, based on the pink bands, is considered in this paper. 

This two-phase halo mass assembly translates into a two-phase baryon mass assembly at
the galactic scale. This is  illustrated in Figure \ref{SFRH}, where we plot 
again M$_{\rm halo}$(t$_{\rm U}$) (black curves) and M$_{\rm bar}$(t$_{\rm U}$),
the baryon mass in the galactic object along the evolution (red curves).
We see that M$_{\rm bar}$(t$_{\rm U}$)  roughly follows M$_{\rm halo}$(t$_{\rm U}$), 
and it also shows a two-phase behaviour, with no major mergers along the slow phase in any of the three panels.
For more details, see for example  Figure 3 in \citet{Obreja:2013},
where the mass aggregation tracks along the main branch of the merger tree (MATs)
 are given at different fixed radii, both for its cold baryons  and stars. 
The M$_{\rm halo}$(t$_{\rm U}$) curves are also plot and a correspondence between the mass increments
(i.e., merger events) at both scales, halo and galaxy, clearly stands out 
\footnote{It is worth noticing that the shapes of these stellar MATs are 
qualitatively consistent with those of  the cumulative stellar mass curves
as a function of the redshift, for
the MW and MW-like galaxies  \citep{vanDokkum:2013,Snaith:2014}.}.

Therefore we can conclude that not only do our simulated galaxies show a two-phase halo mass assembly,
but the baryon mass assembly 
of the whole disk galaxy 
shows a two-phase behaviour too, closely linked to that of its respective halo.

Let us stress that the physical conditions for star formation are very different before and after t$_{\rm vir}$. 
Before virialization,  stellar activity, either star formation (Figure \ref{SFRH}), winds  or SNe explosions,  is very high.
The many fast mergers inject mechanical energy into the early galactic system,
and together with the energy from discrete sources (i.e., energy feedback), they increase the gas turbulence.
At the same time, the systems experiences high rates of infall of low metallicity gas. 
The mixing of infalling with in-situ gas, aided by turbulence, is an effective way to remove metallicity gradients.
In addition, in this violent phase the gravitational potential is time-dependent, and
therefore scattering of the already-formed stellar populations can be expected too.  
Thus, the spatial distribution of stars can change considerably along the fast phase
 relative to that at their birthtime.

In contrast with this situation, during the slow assembly phase the potential at the halo scale is nearly spheroidal and
axial at galactic scales. 
Axial symmetry can occasionally be broken, for example owing to disk disruption caused by   satellites   or when  a  bar develops. 
On the other hand, gas turbulence is expected to be low 
during this phase, because energy injection events are now scarce at any scale.

The expected imprints of  these different $physical$ conditions 
on the properties of the stellar populations born before and after virialization 
are in line with the properties of the thick and thin disk stars, respectively, 
properties highlighted on the basis of an $empirical$ classification.
In section $\S$\ref{ThinThickProp} we show that this is indeed the case.
However,  the $physical$ timescale t$_{vir}$ is not an observable. 
In order to make the comparison to observations possible, 
we need first to find out an $observable$ timescale that is linked to t$_{\rm vir}$ to classify disk stars.
In the next section we show that there is a chemical timescale that matches this requirement.  

\section{Component Classification of Stellar Populations} 
\label{tchem}

To identify  three components in the simulations we proceed in two steps.
First, stars are split into their spheroid
 (bulge and stellar halo) and  disk components,
 using   the  $kk$-means method in the space of   kinematical
variables  \citep[see][]{Domenech:2012,Dhillon:2004}. The following kinematical variables are used:
energy $E$, eccentricity $\frac{j_z}{J_c(E)}$, and $\frac{j_p}{J_c(E)}$,
where $j_z$ and $j_p$ are the projections of a given particle angular momentum
on the disk axis and plane, respectively, and $J_c(E)$ is the angular momentum of the
circular orbit with $E$ energy.
With this method, the spheroid and disk star  properties found in
observations are recovered
\citep{Domenech:2012,Obreja:2013,DominguezTenreiro:2015}.
A similar kinematic selection is made for the GASOLINE galaxy.

In a second step, disk stars have to be assigned to either  the thin or the  thick disk,
based on an empirical timescale linked to t$_{\rm vir}$.  
As said in $\S$\ref{Intro},  the empirical classification methods resting  on a timescale take advantage of the two-slope behaviour of the [$\alpha$/Fe] - age relation, based on solar-neighborhhod stellar data and recently extended to early type galaxies. 
 
The SFRH for the whole galaxies\footnote{We need this complete information because disk stars form out of gas enriched by the ejecta of the explosions and winds of all the co-spatial stars belonging to any component.}
are given in Figure \ref{SFRH}. We see that  early SFRs are high  (i.e., they show a low characteristic timescale for SF) and then they decrease to low values at  later times.
For the  g1536-L$^{\rm *}$ galaxy, the SF timescale contrast between early and late SFRs is less marked
than for the other two galaxies.

The chemical [O/Fe]-age relation for disks  are drawn in Figure \ref{Vir-MAT}, 
where we can see that they show a two-slope behaviour
 (rather steep at high z and much flatter at low z), as  their zeroth order shape\footnote{[Mg/Fe] vs age plots show a two-slope behaviour similar to that of [O/Fe]-age.}.
The slope contrast is less marked for g1536-L$^{\rm *}$ galaxy than for the other two 
(because the SF timescales  contrast between early and late SFRs is less important here),
and in addition it shows a lurch around t$_{\rm U} \sim$ 7  Gyrs, i.e., when its  SFRH (Figure \ref{SFRH})
shows an excess after a dip
\citep[see Figure 10 in][for an explanation]{Snaith:2015}.

This global two-slope behaviour  can be used to define a chemical scale 
separating the thin from the thick disk populations.
The simplest option is to make a two-slope fit to the [O/Fe] - age plot. 
Among the different possible solutions, 
we have chosen the solution t$_{\rm chem}$ that is the closest to t$_{\rm vir}$.
The t$_{\rm chem}$ values are observational proxies for t$_{\rm vir}$. 
Their values for the simulated galaxies are given in Table 1, and are marked as
vertical magenta lines in Figures \ref{Vir-MAT} and \ref{SFRH}.

The next task is to make sure that the stellar populations
classified as belonging to the thick and thin disks (i.e., older and younger than t$_{\rm chem}$ respectively) 
do show different elemental abundance distributions as well as
different kinematics, and indeed different trends in kinematics with age, with  metallicity and with [$\rm\alpha$/Fe],
similar to those found in observational data, see $\S$\ref{Intro}.



\begin{figure*}
\vspace{-10.0cm}
\includegraphics[width=0.99\textwidth]{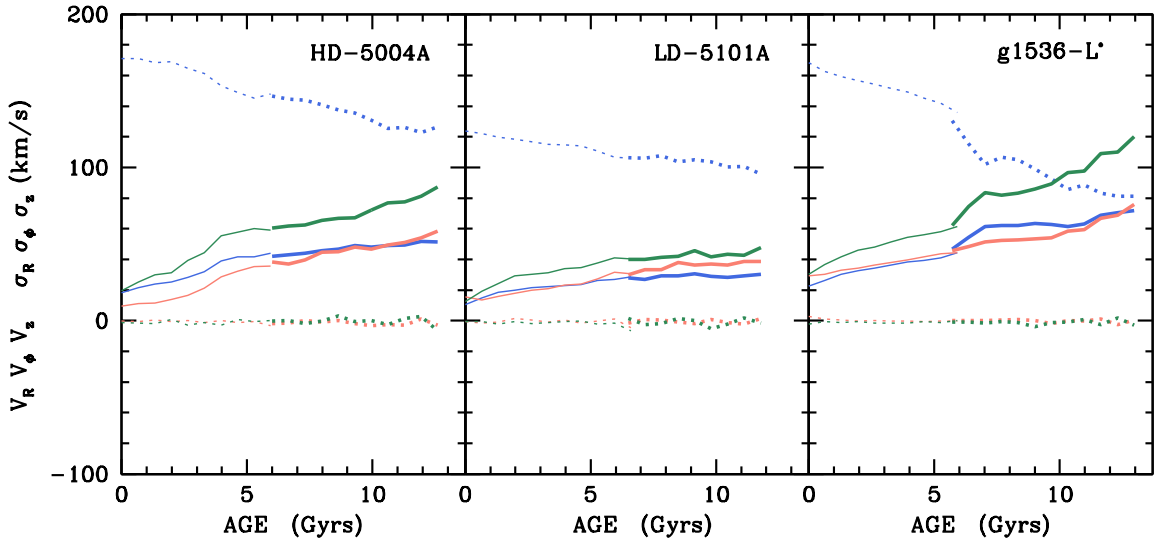}
\caption{Tangential (blue, V$_{\phi}$), vertical (red, V$_{\rm z}$) and radial (green, V$_{\rm R}$) velocities (dotted lines),
and their respective velocity dispersions (solid lines), versus stellar age, for the thin (thin lines) and thick (thick lines) disks analyzed in this work.  
The stars are in a cylindrical shell with 3 kpc $  < R_{\rm cyl} <$20 kpc. 
}
\label{Kine_age_5004_5101}
\end{figure*}


\begin{figure*}
\vspace{-10.0cm}
\includegraphics[width=0.99\textwidth]{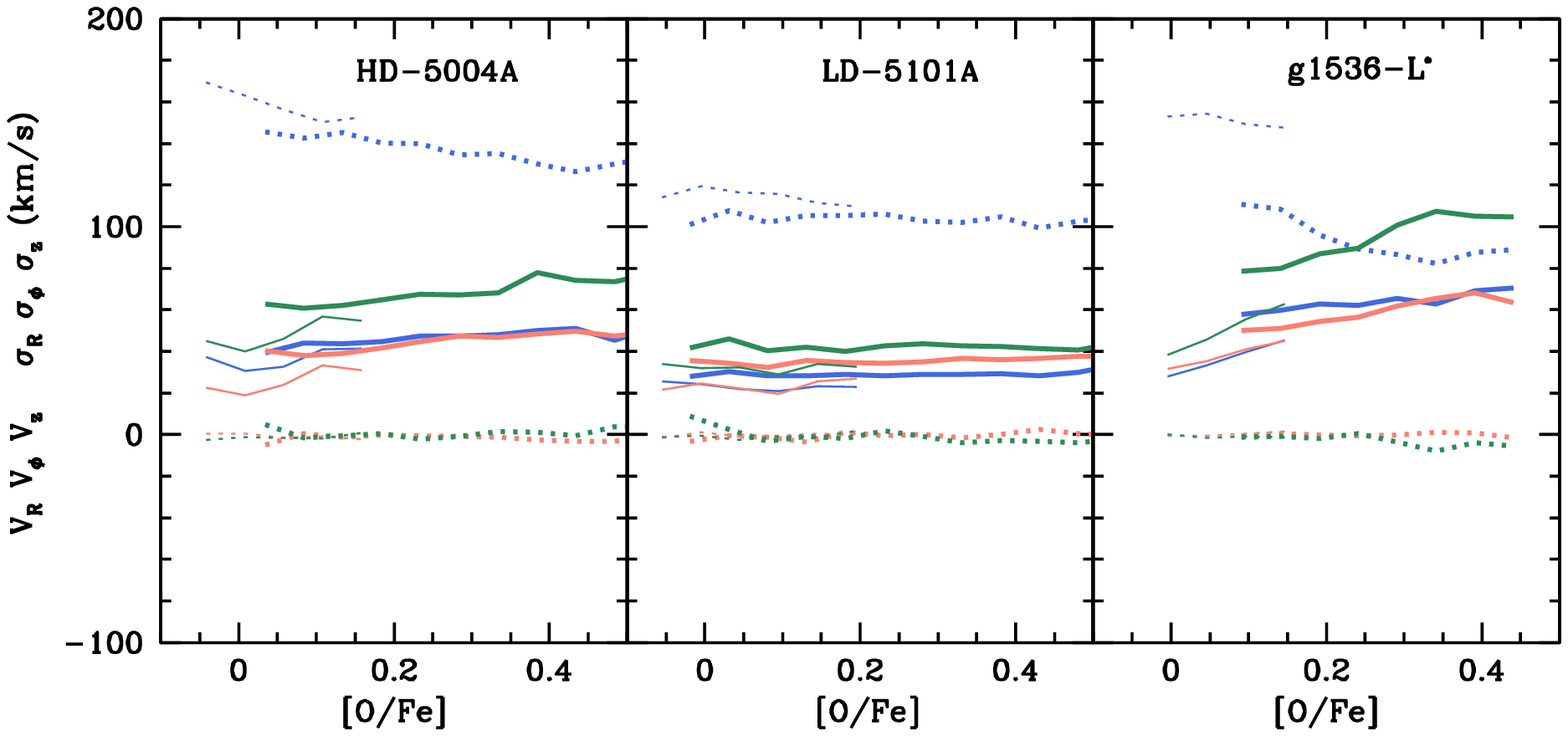}
\caption{Kinematics versus [O/Fe]. Codes  as in Figure \ref{Kine_age_5004_5101}.}
\label{Kine_AlFe_Fe}
\end{figure*} 


\begin{figure*}
\vspace{-10.0cm}
\includegraphics[width=0.99\textwidth]{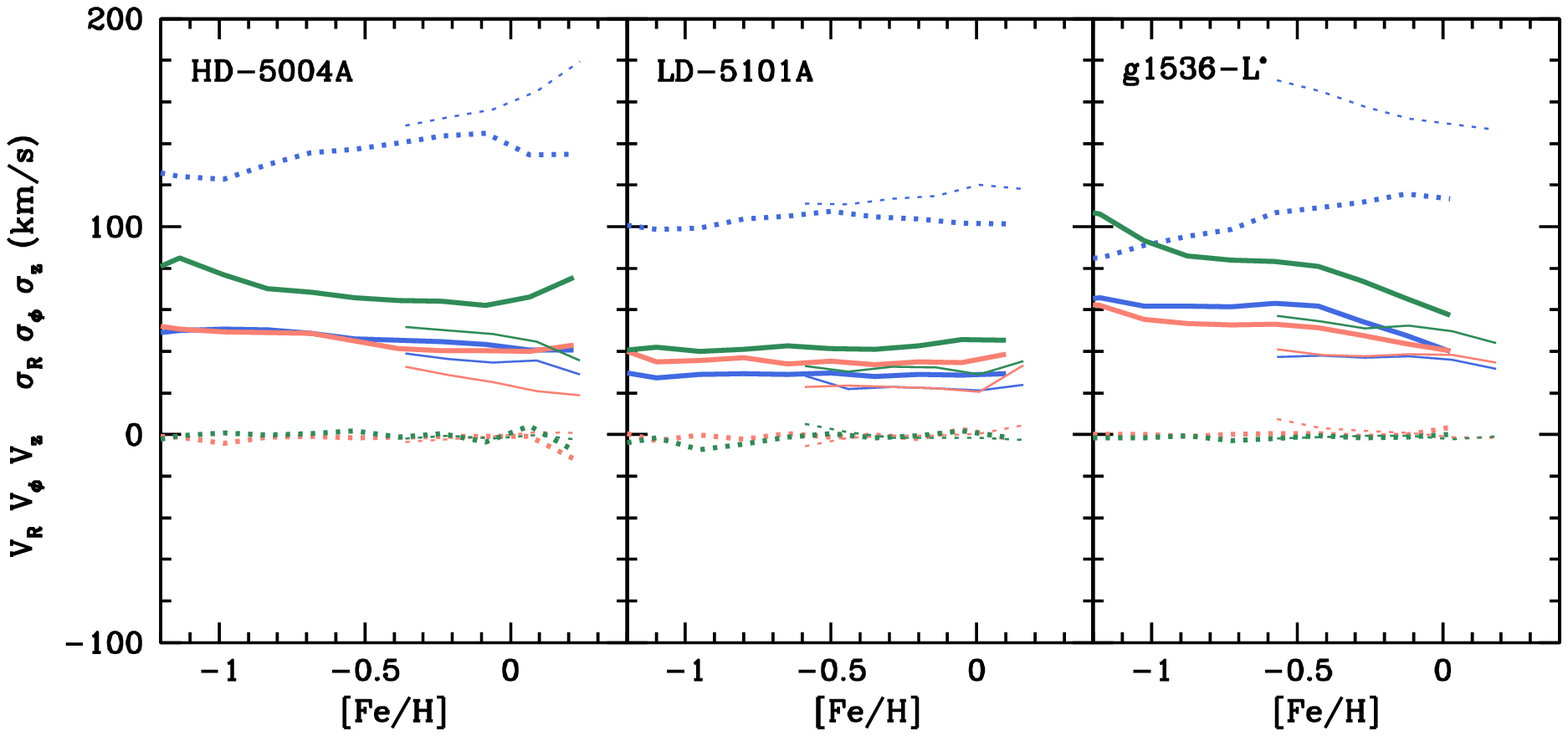}
\caption{Kinematics versus [Fe/H]. Codes  as in Figure \ref{Kine_age_5004_5101}.
}
\label{Kine_Fe}
\end{figure*}


\section{Thin and Thick Disk Properties}
\label{ThinThickProp}

A detailed analysis of the different properties  shown by stars in the thick and thin disk components in a
sample of {\tt P-DEVA} galaxies can be found in \citet{Domenech:2012}.
These authors have analyzed the respective sizes and shapes, as well as the  age, [$\rm\alpha$/Fe] and [Fe/H] distributions
of the different galaxy components according to different classification schemes. Their conclusions are
that the properties shown by thin and thick disk stars in this simulated sample of galaxies are nicely consistent
with observations, irrespective of the classification scheme used.
Similar analyses has been performed  for {\tt GASOLINE} g1536-L$^*$ galaxy \citep[see, for example][]{Stinson:2013b}.

In what follows we go deeper into these different properties, focusing into the different trends in three dimensional kinematics
 with age, metallicity and [$\rm\alpha$/Fe] shown by the disks, and their comparison to observational data.
It is worth to note that detailed data for stellar populations are currently only available for the MW, while
our simulations have been run from random, i.e., non-constrained, initial conditions. Therefore, only 
qualitative consistency between the former and the later can so far be
 required to pass the validation test,
with quantitative agreements adding further strenghths, but not being strictly necessary.

The particle velocity field is usually expressed in cylindrical coordinates
(the symbols V$_{\phi}$, V$_{\rm z}$, and V$_{\rm R}$ will hereafter be used for the tangential, 
vertical, and radial velocities, respectively,
and $\rm\sigma_{\phi}$, $\rm\sigma_{\rm z}$, and $\rm\sigma_{\rm R}$ for their corresponding dispersions).
The velocities are calculated relative to the galaxy center of mass, 
and the system is oriented such that the z-axis at t$_{\rm U}$ is parallel to the angular momentum of the gaseous disk at the same epoch.
All the results shown in this section refer to the 3 kpc $ < $ R$_{\rm cyl}$ $<$  20 kpc disk  shells, 
where R$_{\rm cyl}$ is the radial distance in the cylindrical coordinate system.
Results for the  6 kpc $<$ R$_{\rm cyl}$ $<$ 10 kpc galaxy shells, somewhat similar to the MW solar neighborhood in its 
geometric characterization, show no remarkable differences but a higher statistical noise. 
For this reason, we present only the results in 3 to 20 kpc cylindrical shells.


\subsection{Kinematics versus stellar age}
\label{KineAge}

Figure~\ref{Kine_age_5004_5101} shows the tangential, V$_{\phi}$(age), vertical, V$_{\rm z}$(age),  and radial, V$_{\rm R}$(age),
 velocities (dotted lines) 
and their respective velocity dispersions (solid lines) for the three simulated galaxies. 
Curves corresponding to the thick (thin) disk components are plotted using thick (thin) line types. 
The horizontal axis bin sizes are 0.66 Gyr. No qualitative changes result when using half this value.

We see that in all the cases, as expected, the thin disk appears to be more rotationally supported than the thick one.
The differences at equal R$_{\rm cyl}$ (not drawn) are in the range $\rm\sim$ 35 - 15 km s$^{\rm -1}$, consistent with \citet{Veltz:2008}, 
who find an asymmetric drift of V$_{\rm lag}$ = 33 $\pm$ 2 km s$^{\rm -1}$ for the Galactic thick disk. 
Similar values were found for example by \citet{Chiba:2000} and \citet{Dambis:2009}.
The V$_{\phi}$(age) curves show a decreasing behaviour with increasing age, in consistency with \citet{Haywood:2013},
see their Figure 16\footnote{Note that the yellow points in Figure 16 of \citet{Haywood:2013} belong to the thick disk as defined in this paper.}.
The g1536-L$^{\rm *}$ galaxy shows a remarkable decrement due to major mergers around t$_{\rm U} \sim $ 7 Gyr,
 that changes the ordered radial velocity into velocity dispersion.
Therefore,  our simulated galaxies recover the qualitative behaviour of local stars regarding V$_{\phi}$(age).

The deviations of the V$_{\rm z}$ and V$_{\rm R}$ curves from zero are a measure of the 
misalignments of the stellar disk plane from the cold gaseous one at each t$_{\rm U}$.
No remarkable differences show up, except for weak fluctuations that in some cases become  more important in the old age end, 
where disks are not well defined yet. 
It is worth noticing that these curves inform about stellar disks occurrence as early as 12 Gyr ago.

 Concerning the velocity dispersions, the $\rm\sigma_{\phi}$(age), $\rm\sigma_{\rm z}$(age), and $\rm\sigma_{\rm R}$(age) curves 
show that the thick disk has always higher velocity dispersion values than the thin one.
More specifically, the $\rm\sigma_{\phi}$(age) 
and $\sigma_R$(age)  curves (blue and green solid lines) show up  common
shape patterns  as we go from  young to old stellar populations (patterns not found in  LD-5101A).
After a slow increase with age for young stars, a marked dispersion increase within a short age interval is apparent, followed, again, by a low slope or even flat behaviour.
 Even if the issue is beyond the scope of this paper, we note that these 
g1536-L$^{*}$ and HD-5004A  abrupt velocity dispersion increments occur  
at ages when the disks suffer a particularly violent and/or destabilizing dynamical event. These are: 
 major mergers around t$_{\rm U}$ $\sim$ 7 Gyr (age $\sim$ 6.5 Gyr) for g1536-L$^{*}$, see Figure \ref{Vir-MAT}, 
or a loss of axial symmetry around t$_{\rm U}$ $\sim$ 9.5 Gyr (age $\sim$ 4.0 Gyr) for galaxy HD-5004A, see Figure \ref{Proj_5004}. 
Galaxy LD-5101A is not involved in any such a kind of events, and, consequently, it does not show such remarkable slope changes either.

The vertical velocity dispersion (solid red curves) tends to increase with age too.
However, the changes of slope are less important and less abrupt than the in-plane dispersions.

These patterns are consistent with \citet{Haywood:2013} results, 
see the velocity dispersions curves shown in their Figures 11 and 12, 
where an almost flat behaviour at the young age end, followed by a fast increase, stand out.

Comparing Figure~\ref{Kine_age_5004_5101} with the MW values of these dispersions 
listed in Table 2, we can see that not only the trends but also the numerical values 
of the dispersions are satisfactorily consistent with observational data.
To assess the degree of consistency, 
we should keep in mind that the LD-5101A galaxy is less massive that the MW, 
and that the g1536-L$^{\rm *}$ galaxy has suffered a major merger near t$_{\rm vir}$.
Also, at the old age end (say age older than 10 Gyr) the stellar disk of g1536-L$^{\rm *}$ galaxy is poorly  populated.


\begin{table*}
\begin{minipage}{6.3in}
\begin{center}
\caption{ Milky Way velocity dispersion values from observational data.}
\vspace{0.1cm}
\begin{tabular}{|c|c|c|c|c|} \hline
  &  $\rm\sigma_{\rm R}$ (km s$^{\rm -1}$) & $\rm\sigma_{\phi}$ (km s$^{\rm -1}$) & $\rm\sigma_{\rm z}$ (km s$^{\rm -1}$) & Reference      \\ \hline\hline
Low age end & 25       &   -       & 10       & Haywood et al. 2013 \\
            & 25 $\pm$ 2 & 20 $\pm$ 4 & 10 $\pm$ 1 & Vallenari et al. 2006 \\ \hline
Old age end &    70       &    -      & 45       & Haywood et al. 2013 \\
            & 74 $\pm$ 11& 50 $\pm$ 7 & 38 $\pm$ 7 & Vallenari et al. 2006 \\
            & 63 $\pm$ 6 & 39 $\pm$ 4 & 39 $\pm$ 4 & Soubiran et al. 2003 \\ \hline 
\end{tabular}\\
\label{tab2}
\end{center}
\end{minipage}
\end{table*}



\subsection{Kinematics versus [$\alpha$/Fe]}
\label{KineMgFe}

The kinematics versus [O/Fe] relations are drawn in Figure~\ref{Kine_AlFe_Fe} for the three simulated galaxies,
where the horizontal bin sizes are 0.05 dex. No significant changes occur when using a bin size of 0.025 dex.
We see that the disk tangential velocities, V$_{\phi}$ (dotted blue curves) for thick disks
are always lower than those corresponding to thin disks. They  tend to decrease with 
increasing $\rm\alpha$ enrichment, in consistency with \citet{RecioBlanco:2014} 
 for chemically classified thin-thick disk stars from GAIA-Giraffe data, see their Figure 18.
It is worth noticing that a gap exists between thin and thick populations rotations V$_{\phi}$([$\rm\alpha$/Fe]), even in the [$\rm\alpha$/Fe] ranges where
both populations have similar values, recovering observational trends found by \citet{Haywood:2013} (their Figure 14).

Thick disks vertical and radial dispersion curves,   $\rm\sigma_{\rm z}$  and 
$\rm\sigma_{\rm R}$ respectively, are flat or very slowly increasing  in Figure~\ref{Kine_AlFe_Fe}.
Thin populations curves increase for [O/Fe] higher than $\sim$ 0.0 dex in two galaxies.
Thick disk populations are always dynamically hotter than the corresponding thin ones.

\citet{Lee:2011} have analyzed the kinematics - [$\rm\alpha$/Fe] relations in the SEGUE G-dwarf sample.
By ploting dispersions versus [$\rm\alpha$/Fe] with no thin-thick disk splitting,
they show that the relation is flat at lower [$\rm\alpha$/Fe], and then it increases. 
The same behavior is shown in particular by HD-5004A galaxy
 when no classification of stars is taken into account.
They also found that the dispersions at the low $[\rm\alpha$/Fe] end (mostly young stars, that is,  not affected by specific dynamical events)
increase from  $\rm\sigma_{\rm z}$([$\rm\alpha$/Fe]) to $\rm\sigma_{\phi}$([$\rm\alpha$/Fe]), 
and further to $\rm\sigma_{\rm R}$([$\rm\alpha$/Fe]), 
in consistency with our findings. The numerical values are also consistent for HD-5004A. 
Similar trends have been found in RAVE data \citep{Steinmetz:2006} by \citet{Minchev:2014b}.


\subsection{Kinematics versus [Fe/H]}
\label{KineFeH}

In  Figure~\ref{Kine_Fe} we plot the kinematic curves versus [Fe/H]. We used a horizontal bin size of 0.15 dex and results are stable when e.g. halving this value.
In this figure, the  tangential velocity curves, V$_{\phi}$([Fe/H]), for the thick disks of HD-5004A and g1536-L$^{\rm *}$ (thick dotted blue lines) increase with [Fe/H].
The tangential velocity thin disk curve is clearly decreasing for g1536-L$^{\rm *}$. 
In any case, the thin disk V$_{\phi}$([Fe/H]) curves  show a gap relative to the thick disk ones in the [Fe/H] intervals populated by both thin and thick disk stars.
These patterns of the tangential velocity curves 
are consistent with the findings of \citet{Lee:2011}, their Figure 3;
  \citet{Haywood:2013}, their Figure 14; 
 \citet{RecioBlanco:2014}, their Figure 17;
\citet{AllendePrieto:2016} from GAIA DR1 data combined with APOGEE, their Figure 3.

As for the dispersions in our simulated galaxies,
their shapes for the thick disks are rather flat, except for g1536-L$^{\rm *}$ whose $\rm\sigma_{\phi}$([Fe/H]) (thick solid blue lines) 
and  $\rm\sigma_{\rm R}$([Fe/H]) (thick solid green lines) are slightly decreasing  for [Fe/H] $\rm\gtrsim$ -0.5 dex.
Thin disk dispersions decrease with [Fe/H] for HD-5004A and are rather flat in the other two cases. 
A gap  in the velocity dispersiond between thin and thick populations is found for the three galaxies,
with those corresponding to the thick disk always higher than those corresponding to the thin one.
The highest dispersions are always in the radial direction.
 These behaviours are consistent with  \citet{RecioBlanco:2014} results, see their Figure 20.

To summarize, the well-established fact that thin- and thick-disc stars show different kinematics, 
and indeed different trends in kinematics with stellar age,  metallicity and [$\rm\alpha$/Fe], is nicely recovered  by our simulated galaxies. 
This completes  previous results on thick vs thin disk star properties. 
We therefore have thin and  thick disks in our simulated galaxies, with the correct behavior as expected 
from observations. Thus,  our simulations can be used as a testbed to analyze the origin 
of the thin versus thick disk differentiation, as well as the origin of the MMP radial distributions found by \citet{Bovy:2015}. 
This is the aim of this paper, and the next sections are devoted to study this issue.




\begin{figure*}
\vspace{-2.0cm}
\begin{center}
\includegraphics[width=0.48\textwidth]{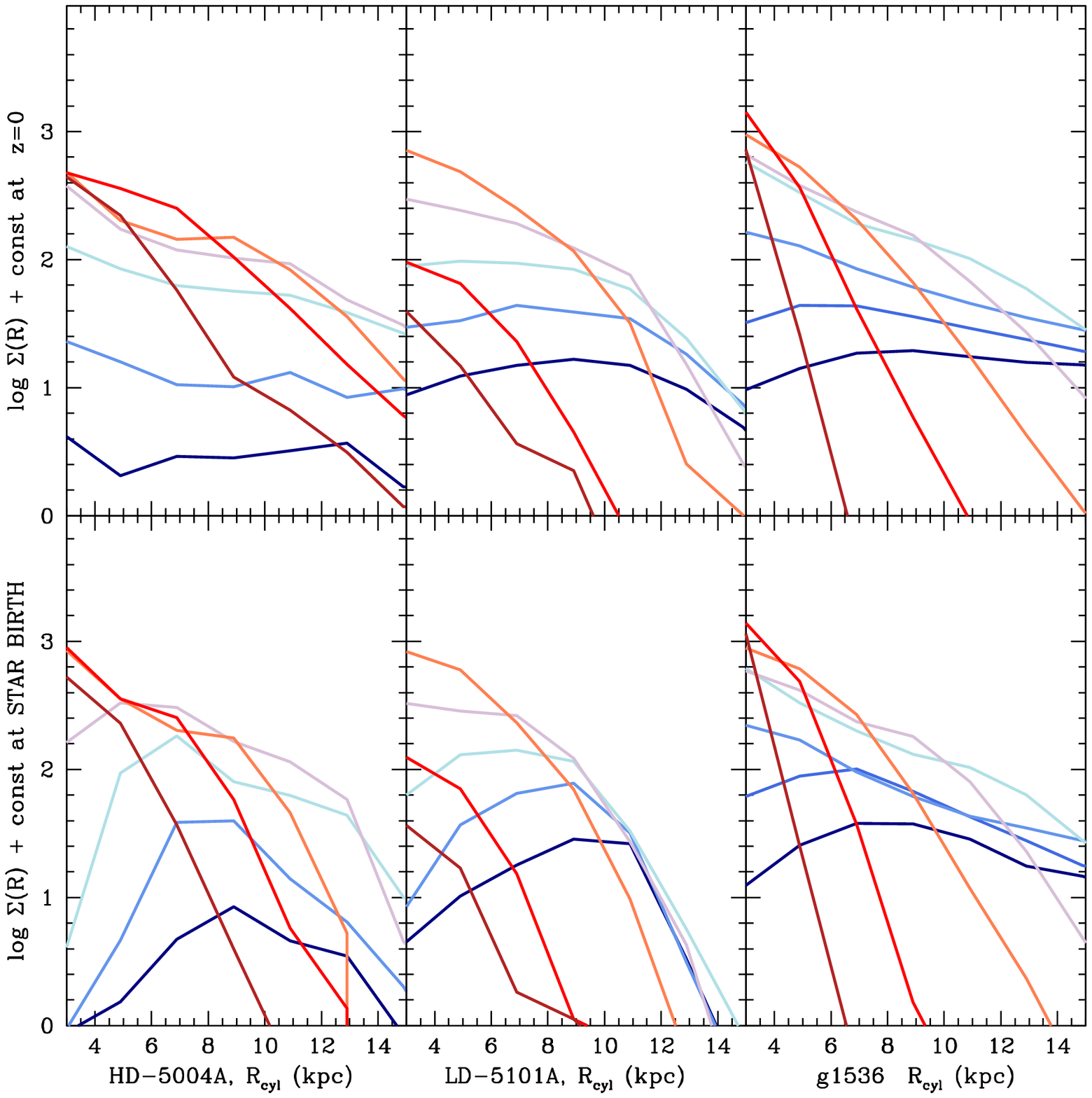} \includegraphics[width=0.48\textwidth]{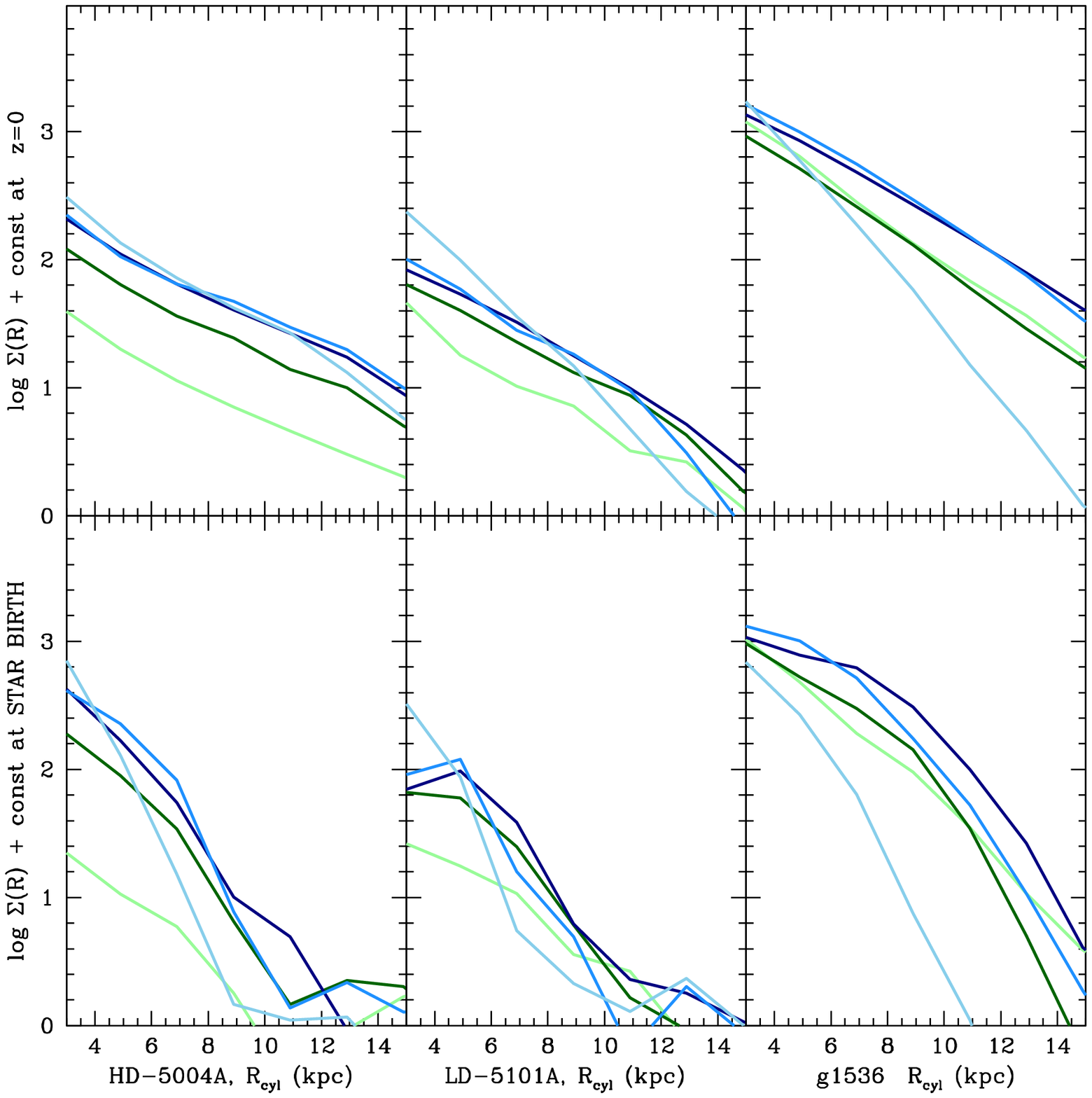}\\
\caption{\textbf{Top row left:} thin disk MMP surface densities for HD-5004A (left), LD-5101A (center) and g1536-L$^*$ (right) galaxies.
Colors code is as follows:
navy blue [Fe/H] $<$ -0.4 and brick color for [Fe/H] $>$ 0.2 ( [Fe/H] $>$ 0.3),
for  HD-5004A and LD-5101A (g1536-L$^*$), the other colors stand for 5 (6) intermediate
[Fe/H] 1 dex wide bins.
\textbf{Bottom row left:} MMP histograms for the  birth sites  of thin disk stars with the same color code as the upper left panels.
\textbf{Top and bottom rows right:} same as top and bottom row left panels, respectively, for the thick disks.
Colors stand for different [Fe/H] bins: pale green [Fe/H] $<$ -0.8, sky blue for [Fe/H] $ >$ -0.25, the other
colors stand for 3 intermediate, $\sim$ 2 dex wide bins.}
\label{MMP-THIN-THICK}
\end{center}
\end{figure*}


\begin{figure*}
\vspace{-2.0cm}
\centering
\includegraphics[width=0.50\textwidth]{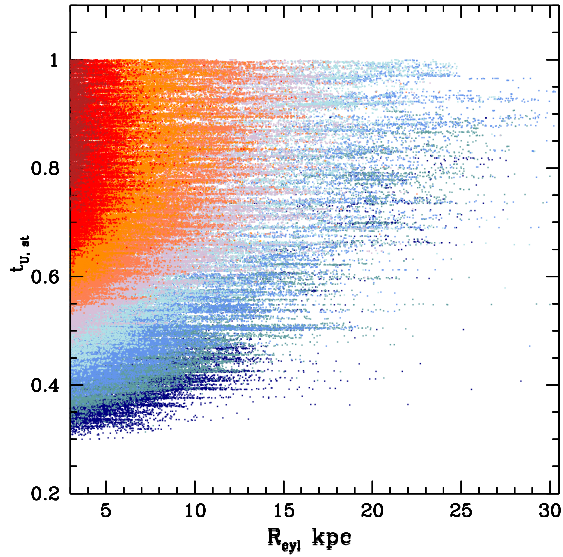}
\vspace{-1.5cm}
\caption{Stellar age in units of the age of the Universe, t$_{\rm U}$, as a function of birth position, R$_{\rm cyl}$, 
for the disk of galaxy g1536-L$^*$.
Colors stand for seven [Fe/H] $\sim$ 0.1 dex wide  bins, plus the low (navy blue, [Fe/H]$<$ -0.5), 
and the high  (brick color, [Fe/H]$>$ 0.2) metallicity  ends.}
  \label{Rcyl-AGE-birth}
\end{figure*}


\begin{figure*}
\centering
\includegraphics[width=0.60\textwidth,height=0.50\textwidth]{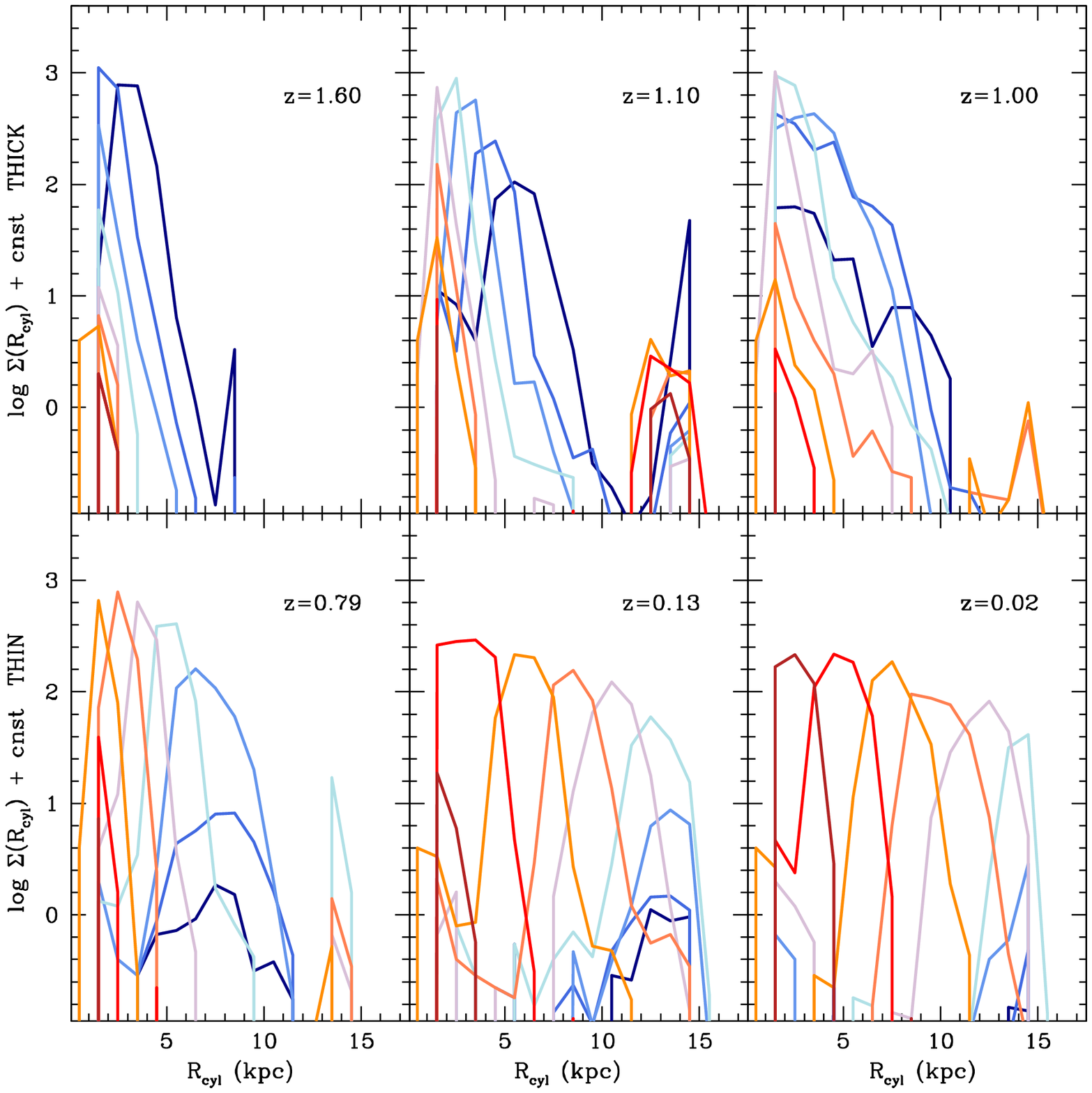}
 \caption{Gas particle R$_{\rm cyl}$ distributions for different [Fe/H] intervals (colors) in HD-5004A galaxy at different redshifts, z.
The top (bottom) row corresponds to the formation period of thick (thin) disk stars. Color code is as in Figure \ref{Rcyl-AGE-birth}}
   \label{Gas-MMP-THIN-THICK}
\end{figure*}


\section{MMP Distributions }
\label{MMPsimu}

The star surface number densities $\rm\Sigma$(R$_{\rm cyl}$) have been calculated for MMPs, and depicted 
in the first row of Figure \ref{MMP-THIN-THICK} for the thin (left) and the thick (right) disk stars, respectively. 
Colors  stand for different [Fe/H] bins, which  for thin disks are
navy blue [Fe/H] $<$ -0.4 and brick color for [Fe/H] $>$ 0.2 ( [Fe/H] $>$ 0.3),
for  HD-5004A and LD-5101A (g1536-L$^*$), the other colors stand for 5 (6) intermediate
1 dex wide [Fe/H] bins. 
For the thick disks, color bins are: pale green [Fe/H] $<$ -0.8, sky blue for [Fe/H] $ >$ -0.25, the other
colors stand for 3 intermediate, $\sim$ 2 dex wide bins.

 The MMP surface number densities show a dichotomy:
i)  the $\rm\Sigma$(R$_{\rm cyl}$) for thick disks have  similar  scalelengths no matter the 
MMP (except for the most metal rich one in g1536-L$^*$), and no breaks \citep[see][for similar results]{Stinson:2013b},
ii)  for the thin disks, $\rm\Sigma$(R$_{\rm cyl}$) have different shapes depending on the MMP:
 those corresponding to high [Fe/H] bins peak at low R$_{\rm cyl}$,
while those corresponding to low [Fe/H] values are rather flat.
Moreover, some slope changes appear that tend to be placed at increasing R$_{\rm cyl}$ for lower [Fe/H]. 
Qualitatively, we recover the APOGEE results for the Galactic MMP radial structure, for high and low [$\rm\alpha$/Fe]
populations (a proxy for old and young populations, respectively).

\section{Star Birth Places and Gas Metallicity Structure}
\label{BirthGasStr}

To decipher the physical processes underlying this behaviour, the birth places for each star belonging to either the thin or the thick disks have been determined. 
In the bottom panels of Figure \ref{MMP-THIN-THICK}  we plot the surface number densities the thin (left) and thick (right) disks would have 
if stars had not moved away from their  birth places.
The [Fe/H] bins are the same as those corresponding to their respective upper panels.
These plots indicate that different MMPs are preferentially born at different locations, characterized by an increasing R$_{\rm cyl}$ 
as the metallicity decreases, the effect being much more marked for the thin than for the thick disk populations. 
We can also see that thick disks had their stars diffused outwards between their birth time and z=0.

Different MMPs are also preferentially born at different epochs, t$_{\rm U}$, as Figure \ref{Rcyl-AGE-birth} shows for the disk of g1536-L$^{\rm *}$, see color code at the caption. 
Note that they are not mono-age populations, see \citet{Minchev:2017} for similar results.
\footnote{The paucity of disk stars born outside the disk seen in Figure \ref{Rcyl-AGE-birth}
is a result of the g1536-L$^*$ galaxy construction. Remember that in this case the halo to be zoomed-in has been chosen to have low merging activity. Results for HD-5004A and LD-5101A show a higher fraction, in consistency with other author´s findings \citep[e.g.,][]{Abadi:2003,Scannapieco:2011,Tissera:2012}.}   
At fixed times, i.e., horizontal cuts in this figure, stars reflect the  gas metallicity structure at their birth.

The  gas distributions at different times, t$_{\rm U}$, are shown in Figure \ref{Gas-MMP-THIN-THICK} for HD-5004A,
within the range of thick and thin disk formation ages (top and bottom rows, respectively). 
Colors stand for different [Fe/H] bins, with the same color code as in  Figure~\ref{Rcyl-AGE-birth}.
In this figure we can see a segregation in galactocentric distance, according to gas metallicity in the thin disk.
This segregation is not removed by metal enrichment due to stellar evolution.
This segregation and its  preservation across time implies that thin disk  stars belonging to different 
MMPs are born at preferential times and disk locations.

Regarding the thick disk, in the early period when its  stars are born,
the gas R$_{\rm cyl}$  segregation due to metallicity is not as important as later on.
In fact, this segregation does not appear until z $\rm\sim$ 1.5, 
and once it appears,  it is almost removed each time a dynamically violent event occurs 
(because they favour metals mixing, see $\S$\ref{TwoPhase}).
These sequences of order-disorder can be appreciated in the top row of Figure \ref{Gas-MMP-THIN-THICK}.
Similar situations arise for the other two simulated galaxies. As a consequence, the stellar MMPs 
differentiation at birth for thick disk stars are not that marked as in the case of the thin disk.

\begin{figure*}
\vspace{-10.0 cm}
 \includegraphics[width=0.99\textwidth]{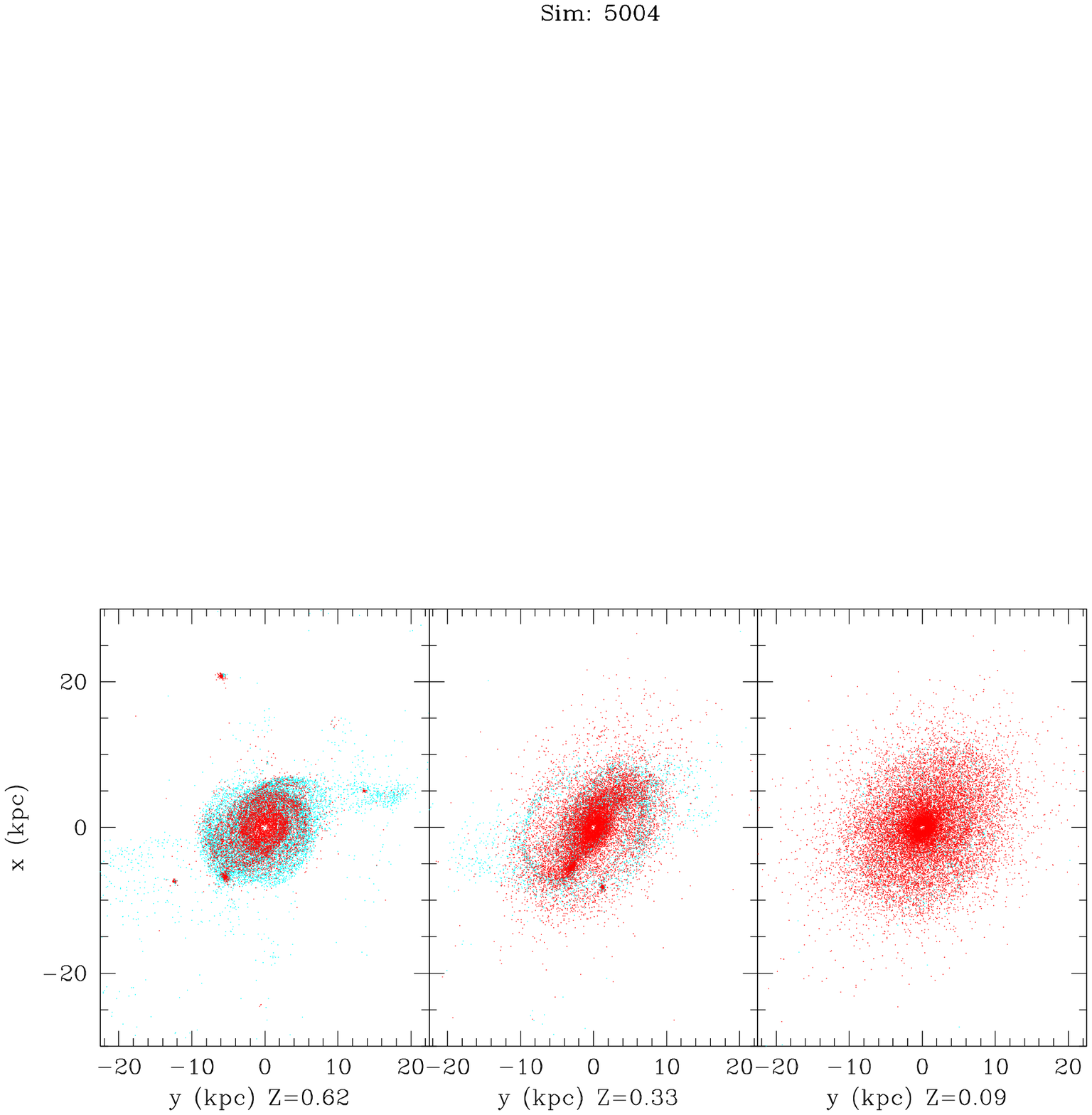}
 \caption{Projection of the baryon particles that at z=0 constitute the thin disk stars of galaxy HD-5004A,
showing its evolution along the slow phase.
Redshifts are written as labels.
Cyan (red) points stand for  gaseous (stellar) particles at the corresponding $z$.
Note the non-axysymmetric structure clearly visible  at z=0.33.}
\label{Proj_5004}
\end{figure*}

\section{Two-phase Mass Assemblies Again}
\label{TwoPhaseAgain}

Violent dynamical events occur much often at early times than later on, along the
thin disk formation period, because of the two-phase characteristic of mass assembly.
This has been  proven and discussed in $\S$\ref{TwoPhase}.
We now focus on how the dichotomy fast / slow rates of mass assembly affect the characteristics
of thin / thick disk stellar populations and, particularly, those of their MMPs radial
distributions.
  
As stated in Sections \ref{TwoPhase} and \ref{tchem}, 
the thin disk begins to form once the galaxy halo reaches its virial equilibrium,
becoming, to a first approximation,  a quasi-stationary system with almost spherical symmetry at the halo scale,
and an axial symmetry at the galaxy scale. 
We have found that, in this situation, 
gaseous disks form inside out and chemical evolution
proceeds within them, building up element gradients as a consequence of stellar evolution.
These gradients show up as the gas R$_{\rm cyl}$  segregation we see in the second row of Figure \ref{Gas-MMP-THIN-THICK}.

In contrast, the thick disk forms its stars while the galaxy halo is still in its fast assembly phase, 
when the halo has a time dependent potential.
In this period, many fast mass increments occur, carrying with them matter 
(dark matter, gas and some stars) either diffuse or clumpy, and energy. 
In this unstable situation, the chemical gradient is built and removed,
as we see in the sequence plotted in the top panels of Figure \ref{Gas-MMP-THIN-THICK}.
Gradients removal occur not only due to the in-falling gas mixing with the local disk gas, but presumably also
helped by an increase in turbulent diffusion caused by the injection of dynamical and stellar feedback 
energy into the system.

Finally we  explore why the R$_{\rm cyl}$ distributions of stellar disk birth places do not match the current distributions at z=0,
either for the thin or the thick disks. To see this, in Figure~\ref{MMP-THIN-THICK} we
compare the first with the second rows on the left (right) panels, for the thin (thick) disks. 
These differences  must be due to stellar radial mixing 
(blurring, radial migration  or scattering\footnote{Blurring refers to stellar orbit epicyclic oscillations around a fixed guiding radius.
Churning or radial migration implies a change in the
guiding center due to an angular momentum change without dynamical heating.
Scattering involves an increase of random energy.}, see e.g. \citet{Sellwood:2002} and \citet{Halle:2015}).

Even if the precise distinction of these mechanisms 
is beyond the scope of this paper, the importance of radial mixing and its correlation with
destabilizing dynamical events is not difficult to asses.

For the thin disk, Figure \ref{MMP-THIN-THICK} indicates that
the radial mixing decreases from HD-5004A to LD-5101A to g1536-L$^{\rm *}$. 
An analysis of the HD-5004A evolution along the slow phase, shown  in Figures \ref{Vir-MAT} and \ref{SFRH}, indicates that 
a burst of star formation occurs around t$_{\rm U}$ $\rm\sim$ 9.5 Gyr.
Within  the same time interval 
 this galaxy develops a non-axial configuration, that stands out at z=0.33 in Figure \ref{Proj_5004}, where we can also see that the destabilizing agent is a double minor merger.
In this way, thin disk stars suffer an important radial migration.

Regarding LD-5101A, no important breaking of axial symmetry appears along the slow phase.
Therefore, radial migration and its effects are   milder than in the previuos case.
The effects of radial migration are even
milder in the case of g1536-L$^{\rm *}$ galaxy. 

The different unstability patterns shown by the three simulated galaxies  along their
slow phase  can be understood in terms of their respective environments. 
Indeed, while HD-5004A lives in a dense environment, LD-5101A lives in a lower density one.
Also, the host halo for the g1536-L$^{\rm *}$ has low merging activity at late times.

These correlations strongly suggest that radial mixing in the slow phase is an effect of disk instabilities, caused by dynamical activity which 
temporarily breaks the axial symmetry of the system. 
Hence, we can say that the final MMP R$_{\rm cyl}$ distributions of the thin disk is shaped by disk instabilities. 
As for the thick disk stars, their formation in a time-varying potential implies a high degree
of dynamical scattering, explaining the uniformity of their R$_{\rm cyl}$ distributions, no matter the MMPs.

\section{Summary, Discussion and Conclusions}
\label{SummDisConc}

It this paper we address the issue of the physical processes underlying the origin of the thick disk relative to the thin disk.
Recent results on radial distributions of mono-metallicity populations (MMPs, i.e., stars in narrow bins in [Fe/H] 
within wider [$\alpha$/Fe] ranges) in the Galactic disk
by \citet{Bovy:2015} had cast doubt on the classical bimodality thin versus thick disk. Our work aims at explaining such 
MMPs radial distributions in terms of these physical processes. We show that these distributions are straightforward consequences of the two-phase
mass assembly scenario for disks.  

It is important to note that in this work 
 similar results have been obtained from the analyses of galaxies simulated using two
very different codes, placing our results on the level of fundamental physical 
processes beyond particular code implementations.

\subsection{Summary of results}
\label{Summary}

\begin{enumerate}
\item Dynamical timescales vs chemical timescales

\begin{itemize}
\item We confirm that halos of simulated disk galaxies have their mass assembled through a two-phase process:
first a fast one, corresponding to halo collapse, when the mass assembly occurs at high rates, and a later one,
with much lower rates, once the halo reaches virial quasi-equilibrium.
\item This two-phase halo assembly causes a two-phase baryon assembly at galaxy scales, either in the gas component
or in the stellar populations. Seen from the assemblying system, a lot of dynamic activity 
(in particular frequent major mergers) occurs along the collapsing
phase, while it is much less important, or even just continuos accretion in some cases, once the halo virializes.
\item Halo virialization  defines a (non-observable) time scale, t$_{\rm vir}$, separating two very different
physical conditions for star formation. Before it, the rates of low metallicity gas infall
are very high,  the gas is expectedly turbulent and
the gravitational potential is time-dependent. 
After it, mass infall is scarce,
star formation occurs in a quiet disk
(except for minor mergers and non-axisymmetric perturbations),
 with a close to axial symmetry, and where the gas turbulence
is expected to be low.
\item 
A generic two-slope behaviour in the [$\alpha$/Fe]-age correlation at early and late t$_{\rm U}$s has been found, providing us with an observable, operationally defined chemical timescale,
t$_{\rm chem}$, an observational  proxy for t$_{\rm vir}$.
 This time, t$_{\rm chem}$, has been  used to classify the stellar populations 
in thick and thin (born before and after t$_{\rm chem}$, respectively). 

\end{itemize}

Summing up, confirming previous findings, we have found that halo mass assembly proceeds
through two well defined phases. 
Its  imprint on chemical evolution at galactic scales has been 
used to classify stellar populations of simulated galaxies into either thick or thin disk populations.
 
\item 
A detailed analysis reveals that the properties of these two populations
recover observational trends of kinematics with age, [Fe/H] and  [$\alpha$/Fe].
Toghether with results from previous analyses by \citet{Domenech:2012} for {\tt P-DEVA} galaxies, 
and by \citet{Stinson:2013b} for g1536-L$^*$ galaxy,
allows us to assert that our simulated
disk galaxies do have thin and thick disk stellar populations, when classified using  chemical
timescales.

\item
The radial structure of MMPs
shows a bimodal behaviour in simulated galaxies, recovering \citet{Bovy:2015} results, see $\S$\ref{Intro}. 

\item
We have found that different MMPs were  preferentially born at different locations, characterized
by an increasing galactocentric distance as [Fe/H] decreases, the effect being much more important
for thin disk stars than for thick disk ones.
Different MMPs were also preferentially born within different Universe age ($t_U$) intervals.

\item
The gas metallicity structure at a given $t_U$ determines the metallicity distributions of stars born at $t_U$.
The radial structure of MMP gas elements has been found to be segregated in metallicity 
after t$_{\rm chem}$ (except for periods of important mixing activity, owing to destabilizing events), 
and rather homogeneous before t$_{\rm chem}$ (due to element
abundance mixing with low metallicity infalling gas, as well as   
to turbulent diffusion).

\item
This structure of star birth places for different MMPs (poorly differentiated for
the thick disk, segregated for thin disk stars) explains the properties of
 the radial distribution of stellar MMPs, $\Sigma($R$_{\rm cyl})$,  at birthtime.
We note the $continuous$ character of the variations of the thin disk $\Sigma($R$_{\rm cyl}$) scale lengths
with metallicity, 
an effect also found
by \citet{Bird:2013} for mono-age populations.

\item  
It has been found that the radial distribution of  MMPs at stellar birth time do not match
those found at z=0. An analysis of the simulated galaxies along
the slow phase indicates that disks can be destabilized, leading to axial symmetry loses that cause radial migration.
It is this late  activity that drives the importance of radial mixing to
determine the final MMP structure in the thin disk. 

Regarding thick disks, the formation of their stars in a time-varying potential (fast phase)  implies a high
degree of dynamical scattering after their formation.
Together with the lack of important gas  metallicity gradients  at their birth times, this   explains
their $\Sigma($R$_{\rm cyl}$) uniformity no matter  the MMP. 

\end{enumerate}

\subsection{Discussion}
\label{Discussion}

Scenarios to explain the thick versus thin disk differentiation have to explain 
two different points: i) when, where, under which physical conditions  and with which properties
did their respective populations form, and
ii) how did these populations attain their current (i.e., z=0) properties, in particular their spatial configurations.
The results summarized above allow us to infer aspects of the thick disk origin.

Basically two different scenarios exist  to explain the thick disk emergence:
 (see $\S$\ref{Intro} and references therein):
 those that link it  to  violent formation processes prior to thin disk formation, from a turbulent gas
\citep[see reviews in][]{Gilmore:1989, Brook:2004}, and those that
assume a preexisting thin disk that is dynamically heated and/or  receives external stellar contributions. 
 By linking the two-component disk concept with the two-phase halo mass assembly
scenario, our results support the first  ones, and in
particular results obtained by \citet{Brook:2004,Brookb:2012}, \citet{Stinson:2013b},
 and  \citet{Bird:2013}, from cosmological simulations.

As said in $\S$\ref{Intro}, different  observational results suggest or are consistent with the previous scenario.
We are  close to the \citet{Haywood:2013}, \citet{Snaith:2014} and \citet{Haywood:2015} proposal,
that the thick disk stars have formed at early times out of a turbulent gas, setting the chemical conditions for a latter thin disk formation in a more quiescent  situation,   
as well as to different results from different surveys (see $\S$\ref{Intro}).

Taking into account the halo environment, cosmological evolution implies a two-phase mass assembly process
because when the cosmological constant becomes dynamically important, a new repulsive force comes into play.
This force causes a slowing down of the merger / mass accretion activity at the scale of galaxy halos
 as Figures 1 and 2 show, \citep[see also, for example][]{Lahav:1991,Lokas:2001,SalvadorSole:2005}. 
A tendency towards freezing-out
of the Cosmic Web evolution also shows up \citep{Robles:2015} because $D^+(t)$, the linear density growth factor, becomes constant at large times. 
From this point of view, a two-phase mass assembly cannot be avoided within the $\Lambda$CDM precision cosmology. 

According to the two-phase scenario, the fast phase involves the collapse of the region surrounding the halo.
Seen from the assemblying system, a lot of merging activity takes place, including stellar satellite
accretions.
We have found that their stellar debris  are incorporated into the host thick disk if
the satellite angular momentum  is approximatelly  parallel to the disk axis (plannar incorporations), 
and into the spheroid otherwise,
as found by \citet{Abadi:2003}, \citet{Meza:2005}, \citet{Scannapieco:2011} and \citet{Tissera:2012},
 see also \citet{DominguezTenreiro:2015}.

This two-phase assembly scenario is consistent with a rather continuous variation of some parameters,
as we have obtained, for example, for the scale lengths of the MMP surface  mass densities in  the thin disks.
Resuts from  other simulations \citep[see, e.g.][]{Bird:2013,Minchev:2017}
or observational works \citep[see, for example][]{Bovy:2012b,Kordopatis:2012,Bovy:2015},
also emphasize this behaviour.
It is worth stressing that it is the distinct physical conditions prevailing at
thick and thin stellar populations formation  that makes the difference between these two components.

The two-phase scenario answers to the question of the generic different physical conditions under
which the stellar populations of the thin and thick disk were born, their ensuing properties,
and  when and where this occurs.
Regarding the question of how did these
populations attain their current configurations, the time-dependent potential
in the fast phase also explains an early scattering of thick disk stars. 
Radial migration in the slow phase could have been a key process   
to match MMP distributions at birthtime to those currently observed in the thin disk.

\subsection{Conclusions}

The occurrence of two distinct phases in halo mass assembly explains the bimodal behaviour  
radial MMPs distributions show according to their [$\alpha/$Fe] content (a proxy for stellar age). 
Indeed, while the frequent dynamical violent events occuring 
at high redshift remove metallicity gradients at early times, and  imply efficient stellar mixing
once the stars are formed, the relatively quiescent dynamics
after the transition keeps [Fe/H] gaseous gradients and prevents newly formed stars to suffer 
from strong radial mixing.

In conclusion, the MMP radial chartography results, as revealed by APOGEE, do not disprove the two-component disk
concept, but they rather reinforce it when analyzed in the wider context of disk formation inside
 a non-isolated halo embedded in its cosmological neighbourhood. 
This new scope reveals that it is the unavoidable emergence of the $\Lambda$-induced force that 
is the underlying physical engine driving the two-phase scenario. And that when an age classification is used,
the stars of the thick disk have formed under completely different physical conditions than those of the thin disk.

To finish, let us remark that, up to now, no significant separating event for the thick and thin disk formation had been identified yet.
Our results indicate that this event is halo virialization, the event marking the transition from the
fast to the slow phase of its mass assembly. This is an important result because it provides  a time scale 
which would even allow predictions to be made from pure Dark Matter  simulations, as well as semi-analytic models, 
linking thick disk/thin disk transition to particular time scales.

\acknowledgments
This work was partially supported through  MINECO/FEDER  AYA2012-31101   and  
AYA2015-63810-P grants (Spain). 
CB thanks the Ramon y Cajal program.
AO thanks the MICINN and MINECO for financial support through a FPI fellowship.
AO has been funded by the Deutsche Forschungsgemeinschaft (DFG, German Research Foundation) -- MO 2979/1-1.
We also acknowledge the Centro de Computaci\'on Cient\'ifica (Universidad Aut\'onoma de Madrid,
Red Espa\~nola de Supercomputaci\'on) for computational support.

\bibliographystyle{apj}

\clearpage

\end{document}